\documentclass[oldversion]{aa}  
\usepackage{graphicx,amsmath}
\usepackage{amssymb}
\usepackage{color}
\usepackage{natbib}		%for A&A : reimplements the LaTeX \cite command 
\usepackage{url}
\bibpunct{(}{)}{;}{a}{}{,} % to follow the A&A style

\def\approxsup{%
  \def\p{%
    \setbox0=\vbox{\hbox{$>$}}%
    \ht0=0.6ex \box0 }%
  \def\s{%
    \vbox{\hbox{$\sim$}}%
  }%
  \mathrel{\raisebox{0.7ex}{%
      \mbox{$\underset{\s}{\p}$}%
    }}%
}

\begin{document}

   \title{Detecting the spin-orbit misalignment of the super-Earth \mbox{55 Cnc e}}

		\author{
		Vincent Bourrier\inst{1}
		\and Guillaume H\'ebrard\inst{1,2}
		}	
	
\authorrunning{Bourrier \& H\'ebrard}
\titlerunning{Detecting the spin-orbit misalignment of the super-Earth \mbox{55 Cnc e}}

\offprints{V.B. (\email{bourrier@iap.fr})}

\institute{
Institut d'astrophysique de Paris, UMR7095 CNRS, Universit\'e Pierre \& Marie Curie, 98bis boulevard Arago, 75014 Paris, France 
\and
Observatoire de Haute-Provence, CNRS/OAMP, 04870 Saint-Michel-l'Observatoire, France}

   \date{} %Received ...; accepted ...}
 
  \abstract
{

We present time-resolved spectroscopy of transits of the super-Earth 55 Cnc e using HARPS-N observations. We devised an empirical correction for the ``color effect'' affecting the radial velocity residuals from the Keplerian fit, which significantly improves their dispersion with respect to the HARPS-N pipeline standard data-reduction. Using our correction, we were able to detect the smallest Rossiter-McLaughlin anomaly amplitude of an exoplanet so far ($\sim$60\,cm/s). The super-Earth \mbox{55 Cnc e} is also the smallest exoplanet with a Rossiter-McLaughlin anomaly detection. We measured the sky-projected obliquity \mbox{$\lambda$ = 72.4$\stackrel{+12.7}{_{-11.5}}^\circ$}, indicating that the planet orbit is prograde, highly misaligned and nearly polar compared to the stellar equator. The entire 55 Cancri system may have been highly tilted by the presence of a stellar companion.   \\

}

\keywords{Planetary systems -- Techniques: radial velocities -- 
     Stars: individual: 55 Cancri}

   \maketitle

\section{Introduction}
\label{intro} 

Spectroscopic observations during the transit of an exoplanet across its host star can measure the sky-projected angle between the spins of the planetary orbit and the stellar rotation (namely the obliquity) through the Rossiter-McLaughlin (RM) effect (\citealt{holt1893}; \citealt{rossiter1924}; \citealt{mclaughlin1924}). The occultation of a rotating star by a planet distorts the apparent stellar line shape by removing the profile part emitted by the hidden portion of the star. This induces anomalous stellar radial velocity variations during the transit, which constrain the sky-projected obliquity ($\lambda$). Whereas first observed systems revealed aligned, prograde orbits (e.g. \citealt{queloz2000}; \citealt{winn2005}; \citealt{loeillet2008}), first misaligned systems were reported with the cases of XO-3 (\citealt{hebrard2008}; \citealt{winn2009a}; \citealt{hirano2011}) and HD\,80606 (\citealt{moutou2009}; \citealt{pont2009}; \citealt{winn2009b}; \citealt{hebrard2010}). About thirty misaligned systems have been identified today over more than eighty measured systems\footnote{\mbox{the Holt-Rossiter-McLaughlin Encyclopaedia:} \mbox{\url{http://www.physics.mcmaster.ca/~rheller/}}} (\citealt{albrecht2012}; \citealt{Crida2014}), including some with retrograde or nearly polar orbits (e.g. \citealt{winn2009c}; \citealt{narita2010}; \citealt{triaud2010}; \citealt{hebrard2011}). These unexpected results favor scenarios where close-in massive planets have been brought in by planet-planet (or planet-star) scattering, Kozai migration, and/or tidal friction rather than more standard scenarios implying disk migration that are expected to conserve the initial alignment between the angular momentums of the disk and of the planetary orbits (see e.g. \citealt{fabrycky2007}; \citealt{guillochon2011}), although some models show that the initial misalignment of a planet can be maintained through its interactions with the disk (\citealt{Teyssandier2013}). Alternatively, it has been proposed that the orbit still reflects the orientation of the disk, with the stellar spin instead having moved away, either early-on through magnetosphere-disk interactions (\citealt{lai2011}) or later through elliptical tidal instability (\citealt{cebron2011}). \\
Obliquity measurements have been mainly done in single-planet systems, mostly on hot-Jupiters. In recent years, they have been extended to transiting multi-planet systems, most of which host super-Earths. Obliquities have been derived from the RM anomaly in the systems KOI-94 (\citealt{Hirano2012}; \citealt{albrecht2013}) and Kepler-25 (\citealt{albrecht2013}), from starspots variations of Kepler-30 (\citealt{SanchisOjeda2012}), and from the measure of the stellar inclination of Kepler-50 and Kepler-60 using asteroseismology (\citealt{Chaplin2013}). These systems have shown coplanar orbits well aligned with the stellar equator, hinting that their orbital planes still trace the primordial alignment of the protoplanetary disk responsible for the planets migration, while the apparent isotropic distribution of obliquities of hot-Jupiters is the result of dynamical interactions (\citealt{albrecht2013}). This conclusion has been recently put in doubt by the large obliquity of the two-planet system Kepler 56 (\citealt{huber2013}) and the possible spin-orbit misalignements of several multi-candidates Kepler systems (\citealt{Hirano2014}; \citealt{Walkowicz2013}).\\
The exoplanet 55\,Cnc\,e offers the opportunity to probe spin-orbit misalignments in the domains of both multiple systems and super-Earths. It is part of a five-planet system, first detected and characterized with radial velocity measurements (\citealt{fischer2008}). The orbital period of the closest and lightest of them, planet 55\,Cnc\,e, was a subject of debates due to aliasing in the radial velocity datasets. The value $P=0.7365$~days proposed by \citet{dawson2010} was confirmed when \citet{winn2011} with MOST, and \citet{Demory2011} with Warm Spitzer, detected photometric transits of planet 'e' at the ephemeris corresponding to that short period. Subsequent studies have refined the orbital and transit parameters of 55 Cnc e using additional photometry and radial velocity measurements (e.g. \citealt{Demory2012}; \citealt{Gillon2012}; \citealt{Endl2012}). 55\,Cnc\,e is thus an unusually close-in ($a=0.015$~au) super-Earth, with a mass $M_p$ = 7.99 $\pm$ 0.25 \,$\mathrm{M}_\oplus$ (\citealt{Nelson2014}) and a radius $R_p$ = 1.99 $\pm$ 0.08 \,$\mathrm{R}_\oplus$ in the optical (\citealt{Dragomir2013}). \\
55\,Cnc is the only naked-eye star hosting a transiting planet. The brightness of that nearby G8V star ($V = 5.95$, $d=12.3$\,pc) makes it a particularly advantageous target for follow-up studies. In particular, it allowed the brightness temperature measurement of 55\,Cnc\,e ($T=2360\pm300$~K) thanks to Spitzer observation of occultations (\citealt{Demory2012}), as well as the possible detection of \ion{H}{i} in the atmosphere of 55\,Cnc\,b ($P=14.652$~days, $M_p$ = 7.8 $\pm$ 0.6 \,$\mathrm{M}_\oplus$) with HST (\citealt{Ehrenreich2012}), indicating that this planet might host an extended atmosphere and suggesting that the orbits of all the planets of the system are nearly coplanar. The orbital evolution of this closely packed system has motivated several studies, which point also toward a coplanar and dynamically stable system (\citealt{Nelson2014}; \citealt{kaib2011}; \citealt{boue2014b}). \citet{kaib2011} showed the 55 Cnc system should be highly misaligned with a true obliquity of $\sim$65$^\circ$, while \citet{boue2014b} pointed out this requires the stellar spin-axis to be weakly coupled to the planets' orbit. Although the detection of the RM anomaly of planet 'e' is expected to be challenging (amplitude $<$ 1m/s), its short period, small radius, and its part in a complex multiple system makes it a particularly interesting target to investigate misalignment. \\
We describe in Sect.~\ref{obs} the observations made with the HARPS-N spectrograph. In Sect.~\ref{correc} we describe the color effect and its correction, in Sect.~\ref{resultsE} we present the detection of the Rossiter-McLaughlin anomaly and in Sect.~\ref{verif} we test its robustness. Discussion of the results will be found in Sect.~\ref{conclu}.

%%%%%%%%%%%%%%%%%%%%%%%%%%%%%%%%%%%%%%%%%%%%%%%%%%%%%%%%%%%%%%%%%%%%%%%%%%%%%%%%%%%%%%%%%%%%%%%%%%%%%%%%%%%%%%%%%%%%%%%%%%

\section{Observations and data reduction}
\label{obs} 

\begin{table*}[th]
%  \centering 
  \caption{Log of the HARPS-N observations.}
  \label{table_log}
\begin{tabular}{lccccccc}
\hline
Run & Transit mid-time (UT) & Exposures$^\dagger$ & Mode$^\ddagger$ & Airmass$^{\dagger\dagger}$ & 
SNR\_392$^\ast$ & SNR\_527$^\ast$ & SNR\_673$^\ast$ \\
\hline
A & 2012-12-26 at 02h54 & 27 & ThAr$^{\ast\ast}$ & $1.10 - 1.00 - 1.35$ & 56 & 261 & 150 \\
B & 2014-01-02 at 01h48 & 33 & FP & $1.00 - 1.43$  & 70 & 342 & 260 \\
C & 2014-01-27 at 02h49 & 30 & FP & $1.00 - 1.38$  & 27 & 144 & 131 \\
D & 2014-02-27 at 01h15 & 30 & ThAr & $1.00 - 1.39$  & 16 & 92 & 93 \\
E & 2014-03-29 at 23h41 & 27 & FP & $1.01 - 1.36$  & 73 & 354 & 269 \\
\hline
\multicolumn{8}{l}{$\dagger$: number of 6-minute individual exposures.} \\
\multicolumn{8}{l}{$\ddagger$: simultaneous thorium-argon (ThAr) or Fabry Perot (FP) reference.}\\
\multicolumn{8}{l}{$\dagger\dagger$: airmass evolution during the observation sequence.} \\ 
\multicolumn{8}{l}{$\ast$: median of the signal-to-noise ratio per pixel at 392\,\AA\ (pipeline order 
\#\,2), 527\,\AA\ (\#\,42), and 673\,\AA\ (\#\,67).} \\ 
\multicolumn{8}{l}{$\ast\ast$: except for the two first exposures made without simultaneous reference.} \\ 
\end{tabular}
\end{table*}

We obtained time to observe a total of eight different transits of 55\,Cnc\,e over three different semesters between late 2012 and early 2014 with the spectrograph HARPS-N at the 3.58-m \emph{Telescopio Nazionale Galileo} (TNG, La Palma, Spain). HARPS-N is a fiber-fed, cross-dispersed, environmentally stabilized echelle spectrograph dedicated to high-precision radial velocity measurements (\citealt{cosentino2012}). It provides the resolution power $\lambda/\Delta\lambda=115\,000$. The light is dispersed on 69 spectral orders from 383 to 690 nm. Due to weather and technical issues, three of the eight scheduled transits could not be observed at all. The log of the five observed transits (runs hereafter labelled from A to E) is reported in Table~\ref{table_log}. Runs A to D were executed in service mode by the TNG Team whereas Run E was made in visiting mode by us.\\
All the observations were sequences of about thirty successive exposures of 6-minute duration each. We chose that duration as a compromise between accuracy, temporal resolution, and overheads. Each sequence lasts several hours (about three hours typically, whereas the full transit last 1.5~hour).
Due to technical reasons, a poor coverage of the transit was obtained during Run A, with only 6 measurements secured during the transit itself. The four other runs allowed a good coverage of the whole transit duration to be obtained. Reference observations were secured immediately before and after the transit for Runs B, D, and E; for Runs A and C, those reference observations were secured mainly after the transit. The observations of Run E had to be stopped earlier than scheduled due to a sudden degradation of the weather conditions after the end of the transit.\\
The CCD was used in its fast readout mode with a speed of 500\,kHz. We used the two 1''-wide optical-fiber apertures: the first one was on the target whereas the second one was used for simultaneous radial-velocity reference, using the thorium-argon lamp or the Fabry Perot depending of the run (see Table~\ref{table_log}). The two first exposures of Run~A are an exception, as they were observed without simultaneous reference, the second aperture being on the nearby~sky.\\
Due to different weather conditions (seeing and absorption), the signal-to-noise ratios (SNRs) were different among the five runs. The Table~\ref{table_log} reports typical SNRs for each run in three different parts of the spectra: Runs C and D were obtained in poor conditions and provide data of relatively low accuracy, Runs B and E were obtained in good conditions and provide particularly high-accuracy data, whereas Run A is intermediate.\\
The SNR values reported in Table~\ref{table_log} are the median among the $\sim30$ exposures of a given run in a given spectral order. In fact, the SNR is varying significantly with time during a given run, and these variations are of different amplitudes from one spectral order to the other. This means that there is a global variation of the flux during a run, but also a variation of the repartition of the flux with the color (hereafter named ``color effect"). The global and chromatic SNR variations show random structures at different time scales, which are probably mainly due to short-term variations of the weather conditions. They also show regular, lower-frequency variations which could be explained by the airmass change of the target during each run, which translates into wavelength-dependent throughput variations. The airmass evolution during each run is reported in Table~\ref{table_log}. The airmass monotonously varies for all the runs but the first one. In the case of Run~A the target reached the meridian during the sequence; it implied an interruption of the observations after the seven first exposures, then a change of the orientation of the alt-azimuth TNG telescope before starting again the observations. The few radial velocities obtained during the transit A were thus secured in different conditions than the reference ones obtained after that transit. Runs B to E were fully executed after the target reached the meridian.\\
The HARPS-N spectra were extracted from the detector images with the DRS pipeline, which includes localization of the spectral orders on the 2D-images, optimal order extraction, cosmic-ray rejection, corrections of flat-field, wavelength calibration with thorium-argon lamp exposures made during the afternoon, and short-term radial-velocity drift correction from simultaneous references with thorium-argon or Fabry Perot. Then the spectra passed through weighted cross-correlation with a G2-type numerical masks following the method described by \citet{baranne1996} and \citet{pepe2002}. All the exposures provide a well-defined, single peak in the cross-correlation function (CCF), whose Gaussian fits allow the radial velocities to be measured together with their associated uncertainties. We tested different kinds of numerical masks as well as removing some low-SNR spectral orders from the cross-correlation; this did not significantly change the observed radial velocity variations. All these procedures were made for Run A using the version 3.6 of the the HARPS-N DRS pipeline, which did not include any correction of the color effect. The DRS version 3.7 which includes a correction of the color effect was available in 2014, and we used it for the data of Runs B to E.

%%%%%%%%%%%%%%%%%%%%%%%%%%%%%%%%%%%%%%%%%%%%%%%%%%%%%%%%%%%%%%%%%%%%%%%%%%%%%%%%%%%%%%%%%%%%%%%%%%%%%%%%%%%%%%%%%%%%%%%%%%
\section{Empirical correction of the color effect on the radial velocities}
\label{correc}

\begin{figure}[]	% gauche , bas, droite , haut
\includegraphics[trim=0cm 2.4cm 2cm 11cm, clip=true,width=\columnwidth]{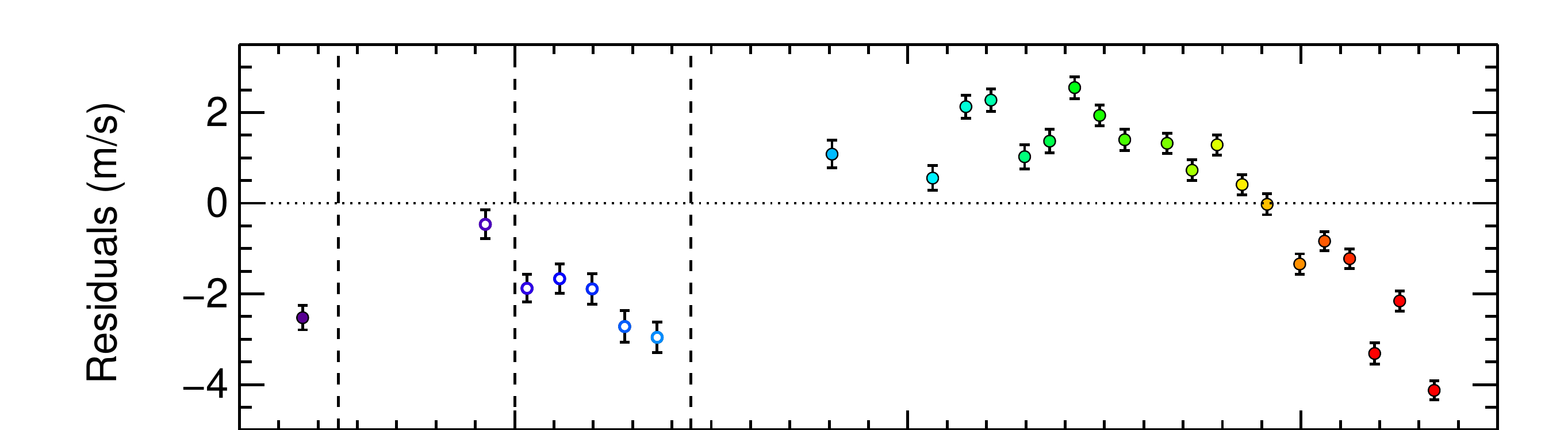}	
\includegraphics[trim=0cm 0.5cm 2cm 11.5cm, clip=true,width=\columnwidth]{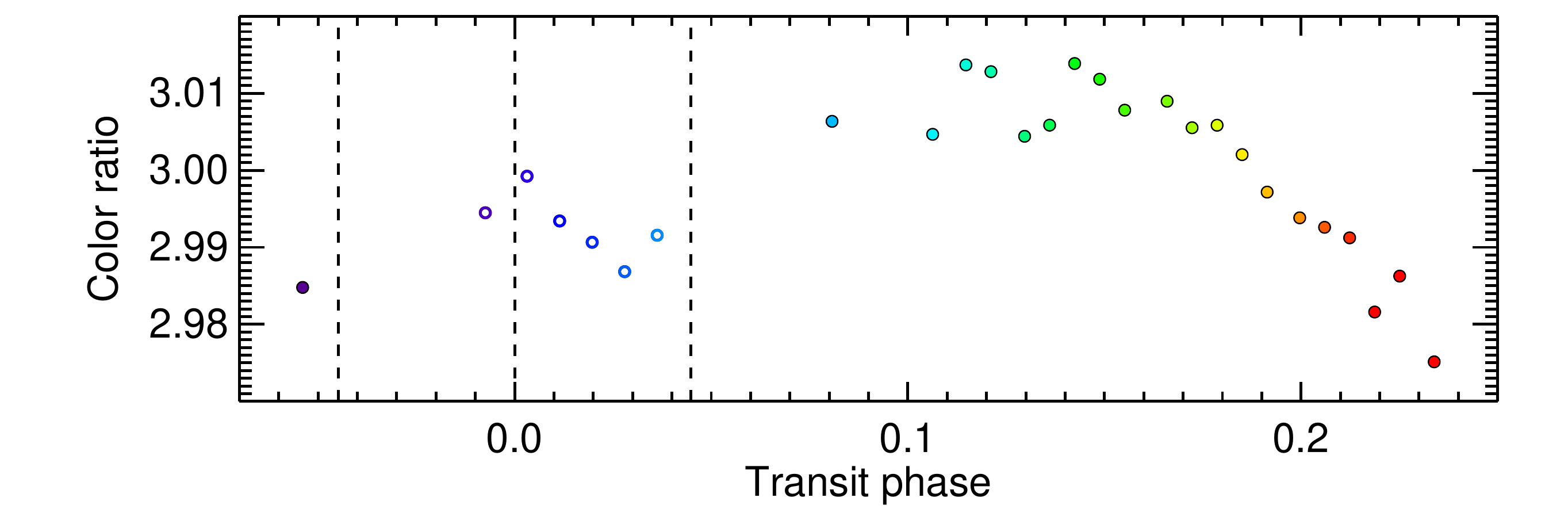}
\includegraphics[trim=0cm 0cm 2cm 11.5cm, clip=true,width=\columnwidth]{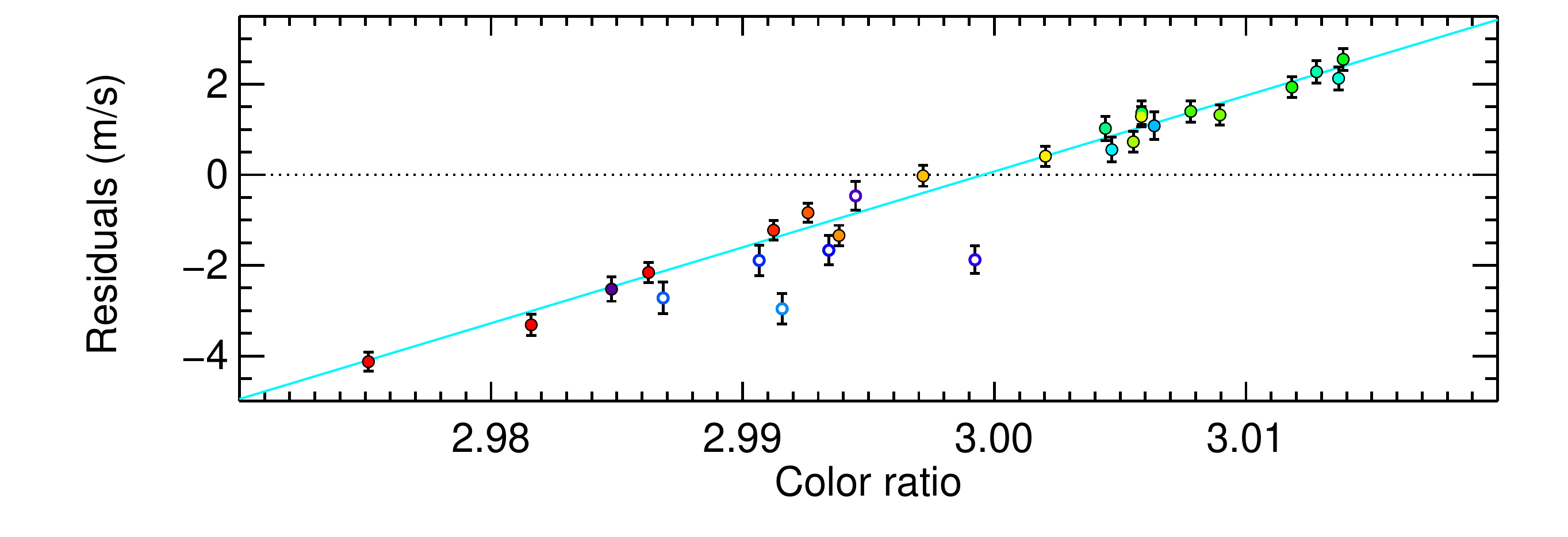}
\caption[]{\textit{Top:} Residuals from the Keplerian fit in dataset A, as a function of orbital phase. Vertical dashed lines show the times of mid-transit, first, and fourth contacts. The colors of the plotted circles indicate the orbital phases of each observation. \textit{Middle:} Color ratio $C_\mathrm{32,32}^\mathrm{28,30}$ as a function of orbital phase. There is a clear correlation with the RV residuals. The decrease of the color ratio at the end of the sequence is mainly due to the increase of the airmass. \textit{Bottom:} Linear relation between the residuals of the Keplerian fit and the color ratio. The fit is performed on the measurements outside of the transit (filled circles); those in the transit (empty circles) roughly follow the same trend.}
\label{fig:runA}
\end{figure}

\begin{figure}[]
\includegraphics[trim=0cm 10cm 2cm 2.8cm, clip=true,width=\columnwidth]{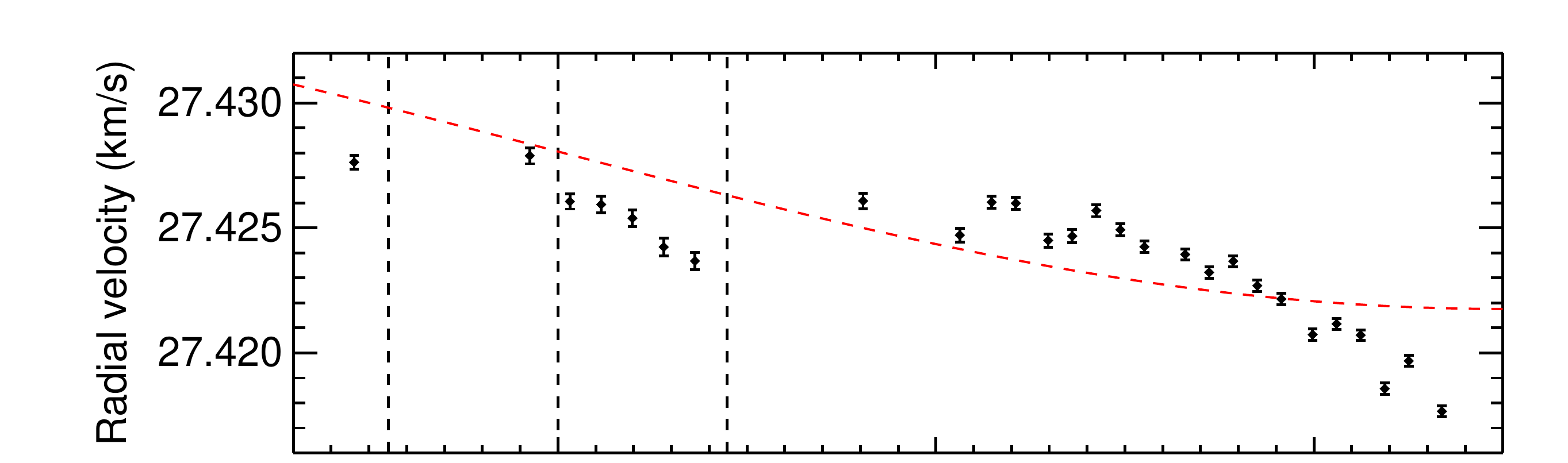}		% gauche , bas, droite , haut
\includegraphics[trim=0cm 8cm  2cm 3.5cm, clip=true,width=\columnwidth]{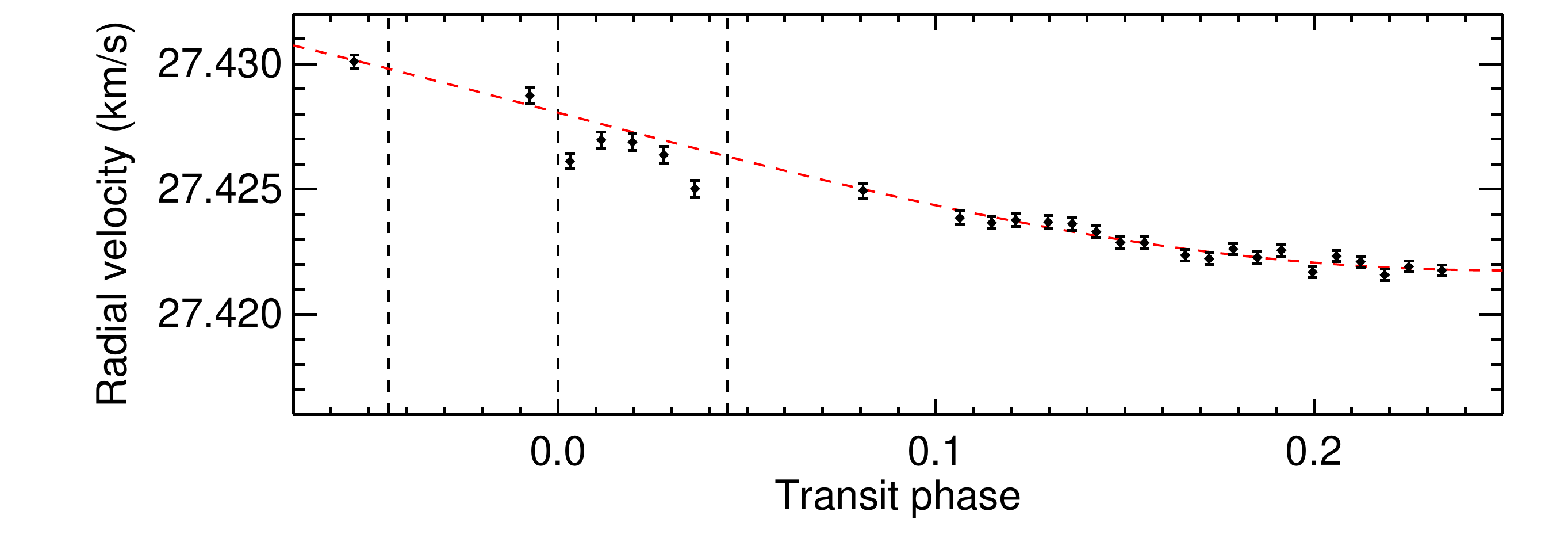}
\caption[]{\textit{Top:} Radial velocity measurements without any color-effect correction (black points) and their Keplerian fit (dashed red line) during Run A. Vertical dashed lines show the times of mid-transit, first, and fourth contacts. \textit{Bottom:} After the empirical color-effect correction, radial velocity measurements outside of the transit are well-adjusted to the Keplerian fit, improving the out-of transit dispersion from 1.92\,m/s to 0.23\,m/s.}
\label{fig:runA_kepler}
\end{figure}

\begin{figure}[]	% gauche , bas, droite , haut
\includegraphics[trim=0cm 2.4cm 2cm 11cm, clip=true,width=\columnwidth]{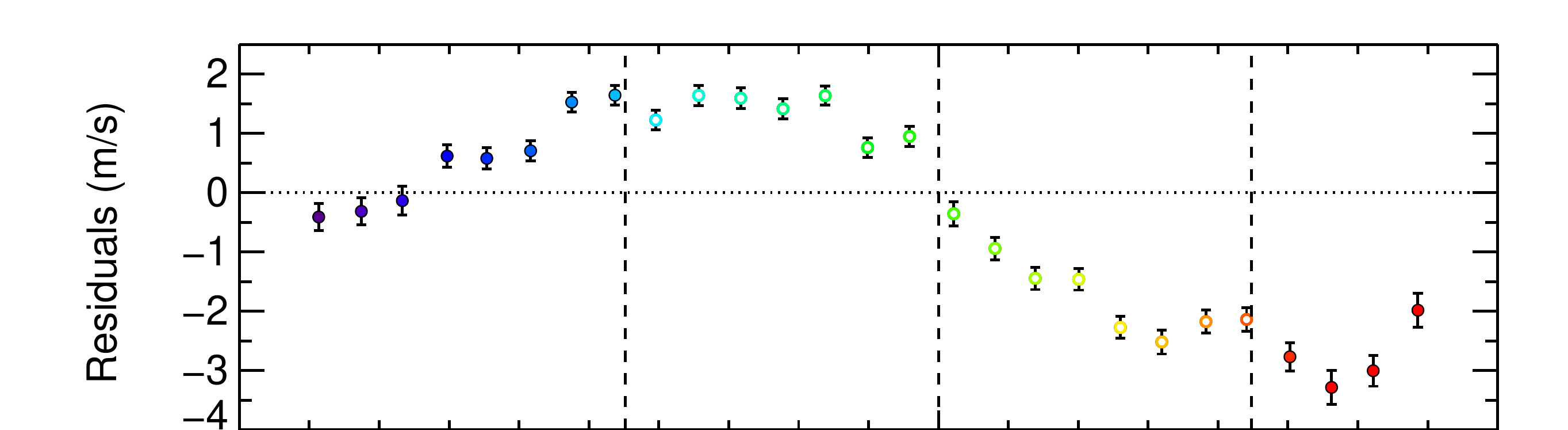}	
\includegraphics[trim=0cm 0.5cm 2cm 11.5cm, clip=true,width=\columnwidth]{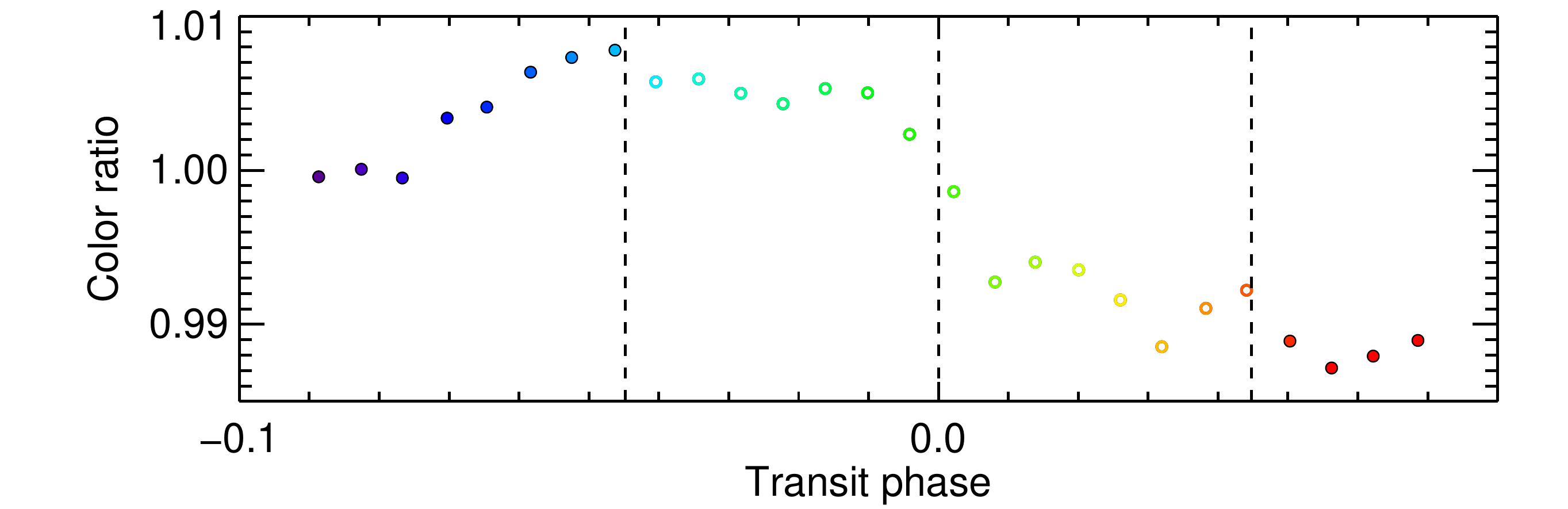}
\includegraphics[trim=0cm 0cm 2cm 11.5cm, clip=true,width=\columnwidth]{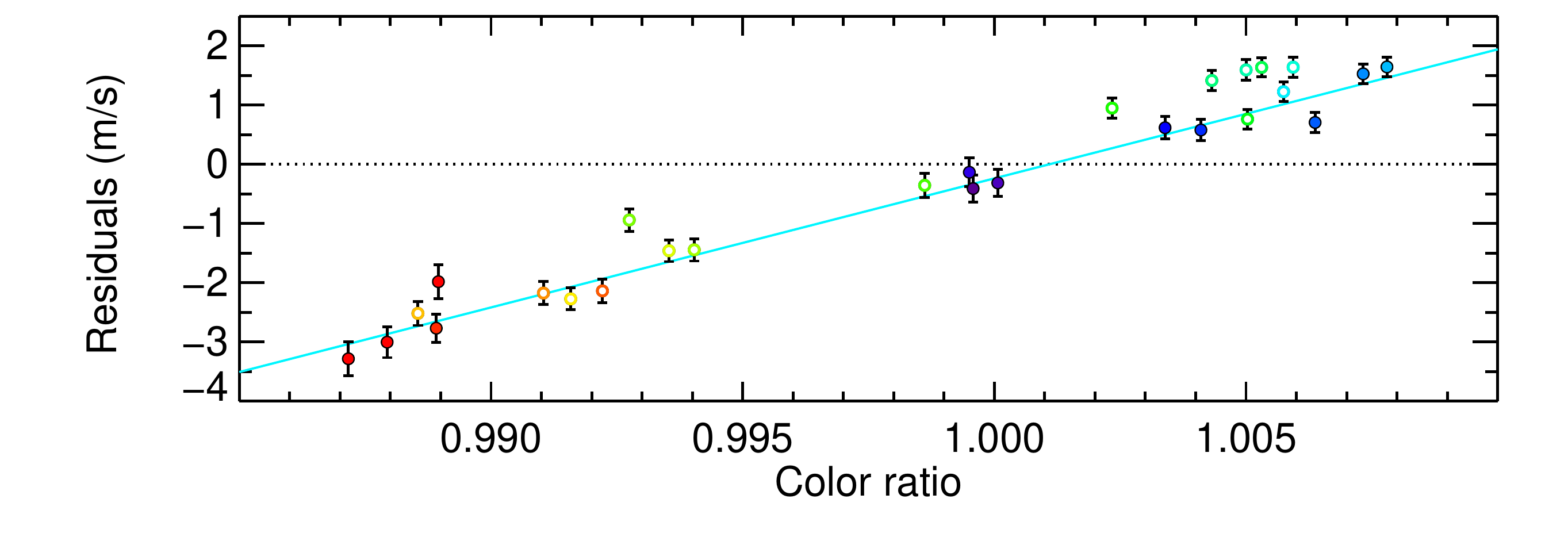}
\caption[]{Same plot as in Fig.~\ref{fig:runA} for dataset E. Again, there is a linear correlation between the RV residuals and a color ratio, in this case $C_\mathrm{28,28}^\mathrm{21,21}$. As for Run A, the fit is performed on the measurements outside of the transit (filled circles) but those in the transit (empty circles) roughly follow the same trend.}
\label{fig:runE}
\end{figure}

The radial velocities of 55 Cnc were fitted with a Keplerian model taking into account the five planets of the system. For each run the fit is performed on the measurements outside of the transit of planet 'e', assumed to be on a circular orbit; its transit epoch, transit duration, and period were taken from \citet{Dragomir2013}, and the semi-amplitude of its radial velocity variations from \citet{Endl2012}. Parameters for the other planets were also taken from \citet{Endl2012}. \\
We observed a trend over each entire run in the radial velocity residuals from the Keplerian fit (see example for Run A in Fig.~\ref{fig:runA}), which we interpret as being due to the color effect (Sect.~\ref{obs}). The CCF represents a mean profile of the thousands of stellar lines in the 69 HARPS-N spectral orders, whose Gaussian fit provides the radial velocity measurement. Because the flux color balance between the spectra varies during a run, this impacts the relative contribution of each spectral order to the mean Doppler shift of the CCF, and thus the measured radial velocity. \\
To characterize and quantify the chromatic variations, we defined a ``color ratio'' between the SNRs associated to the 69 spectral orders:\\
\begin{equation}
C_\mathrm{j1,j2}^\mathrm{i1,i2}(\phi)=\frac{\sum\limits_{i1}^{i2} SNR_\mathrm{i}(\phi)}{\sum\limits_{j1}^{j2} SNR_\mathrm{j}(\phi)}.  
\label{eq:color_ratio}   			
\end{equation}
where $SNR_\mathrm{k}(\phi)$ represents the SNR of spectral order $k$ ($k$ varying from 0 to 68, following the DRS pipeline orders numbering) at orbital phase $\phi$. SNRs at the numerator are summed between spectral orders $i1$ and $i2$ (included), SNRs at the denominator are summed between $j1$ and $j2$. We looked for the combination of spectral orders that gives the best correlation between the color ratio and the RV residuals to the Keplerian fit. For each combination we fitted the data outside of the transit with a polynomial regression, using the Bayesian information criterion (BIC) to prevent over-fitting with a high-order polynomial (\citealt{Crossfield2012}; \citealt{Cowan2012}). For each run we found a good correlation between the variations of the color ratio and those of the residuals to the Keplerian fit. As an example, Fig.~\ref{fig:runA} shows the similarities between the variations of these two quantities in the case of Run A, as well as their linear correlation. This allows us to apply an empirical correction of the color effect to the radial velocities, which improves their adjustment to the Keplerian curve (see e.g. Run A in Fig.~\ref{fig:runA_kepler}). The correlation is also shown in Fig.~\ref{fig:runE} in the case of Run E. As explained below, this run has the best transit observation. \\

%%%%%%%%%%%%%%%%%%%%%%%%%%%%%%%%%%%%%%%%%%%%%%%%%%%%%%%

Runs B to E include a DRS standard color-effect correction which was not available for Run A (see Sect.~\ref{obs}). We could thus apply our empirical correction on datasets B to E extracted without the DRS standard color-effect correction, and compare the two different methods. Table~\ref{table:correc} shows the dispersions of the residuals to the Keplerian fit outside of the transit for both color-effect corrections. The dispersion is always smaller in the case of our empirical correction, in some cases decreasing by more than a factor 2. With the present observations of 55 Cnc e, the empirical correction thus appears to give a better correction of the color effect than the DRS standard correction (Fig.~\ref{fig:residus} shows dataset E reduced with both methods), and we adopt it hereafter. \\

We show in Table~\ref{table:correc} the best-fit parameters for the empirical correction of each dataset. The best correlation between color ratio and RV residuals is always linear, except for run B which requires a 3$^{rd}$ order polynomial correction. Although this dataset has a high precision, it is apparently affected by additional systematics and shows oscillations with an amplitude up to several dm/s. We did not find any correlation between these oscillations and the color effect or any other parameter, and their origin is unclear. Datasets C and D have low SNRs and still show a large, uncorrelated dispersion after correction (the last measurement of dataset D has to be excluded to find an acceptable correlation between color ratio and RV residuals). Dataset A was obtained in different conditions than the other datasets, in particular with data during the transit secured in different conditions than the reference ones outside the transit (see Sect.~\ref{obs}). Run E thus provides the best dataset, with a good sampling and a RV residuals dispersion after our empirical correction which is improved to the level where the Rossiter-McLaughlin anomaly can be detected for a Super-Earth such as 55 Cnc e (Fig.~\ref{fig:residus}). For these reasons, we first fit the Rossiter-McLaughlin anomaly in dataset E only (Sect.~\ref{resultsE}), and analyse in a second time the influence of the other datasets (Sect.~\ref{verif}).

\begin{table*}[tbh]\centering
\caption{Best parameters for the empirical correction of the color effect in each run.}
\label{table:correc}
%\begin{minipage}[b]{0.9\textwidth}
\begin{tabular}{l|ccc|c|c}
\hline
\noalign{\smallskip}  
				& \multicolumn{3}{c|}{\textit{Empirical correction}}		    &	\textit{DRS standard correction}		   	 & \textit{Without correction}										\\   
\noalign{\smallskip}
\hline
\noalign{\smallskip}
 Run    & Color ratio  		& Polynomial		& Dispersion 						&  Dispersion							&  Dispersion  \\  
        & [i1,i2]/[j1,j2]	& degree    		& 	(m/s)								& (m/s)	 								&  (m/s)	 			           \\ 
\noalign{\smallskip}
\hline    
\noalign{\smallskip}
A 			& [28,30]/[32,32]  & 1						&  0.23 										&	-		  							&  1.92			 							     \\  % [0.2cm]
B 			& [15,17]/[12,12]  & 3						&  0.32											&	 0.73								&  2.57		 							\\    %  [0.2cm]
C 			& [14,16]/[13,13]  & 1				  	&  0.67											&	 0.95								&  1.11		 						\\    %  [0.2cm]
D 			& [6,7]/[4,5]  		 & 1						& 0.60											&	 1.52								&  1.39		 							\\  %   [0.2cm]
E 			& [21,21]/[28,28]  & 1						& 0.28											&	 0.43								&  1.86  	 							\\ 
\noalign{\smallskip}
\hline
\noalign{\smallskip}
\multicolumn{6}{l}{Notes: Dispersion is calculated outside of the transit on the residuals to the Keplerian fit.}\\
\multicolumn{6}{l}{The corrections refer to the different color-effect corrections on the radial velocities.}\\
\end{tabular}
%\end{minipage}
\end{table*}

%%%%%%%%%%%%%%%%%%%%%%%%%%%%%%%%%%%%%%%%%%%%%%%%%%%%%%%%%%%%%%%%%%%%%%%%%%%%%%%%%%%%%%%%%%%%%%%%%%%%%%%%%%%%%%%%%%%%%%%%%%

\section{Detection of the Rossiter-McLaughlin anomaly and obliquity measurement}
\label{resultsE} 

\begin{figure}[]   % gauche , bas, droite , haut
\includegraphics[trim=0cm 10cm 2cm 3cm, clip=true,width=\columnwidth]{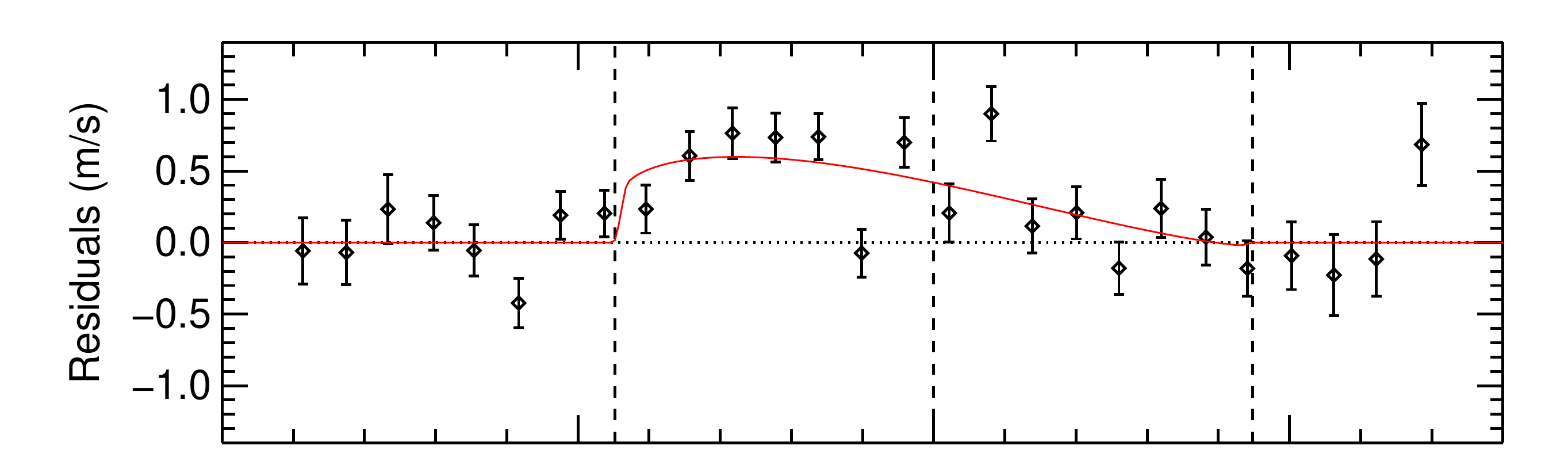}
\includegraphics[trim=1.5cm 2cm  2.08cm 11.9cm, clip=true,width=\columnwidth]{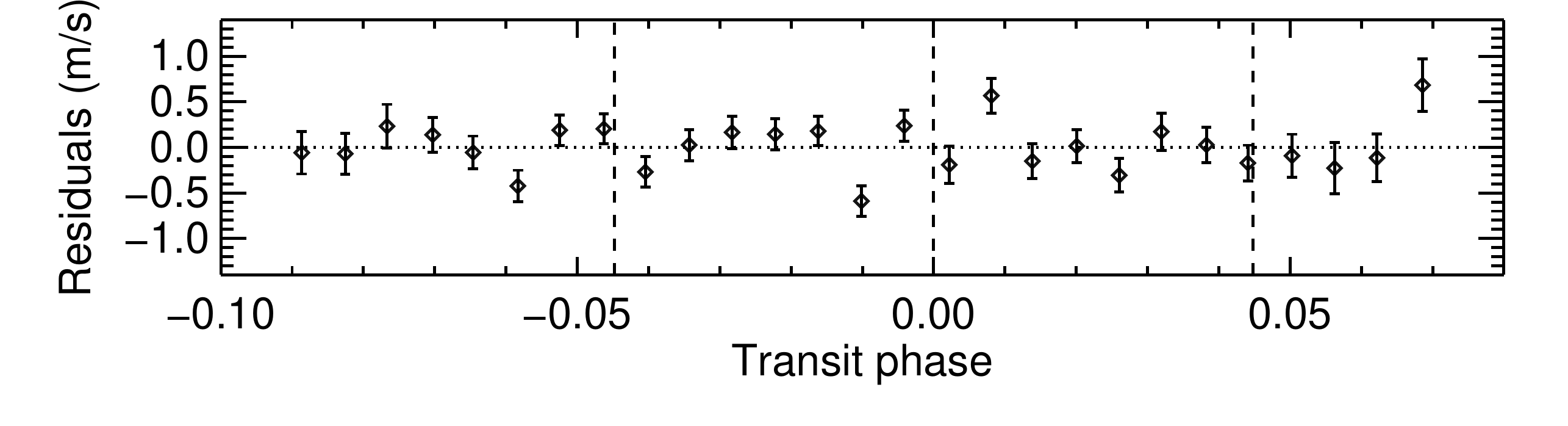}
\includegraphics[trim=0cm 10cm 2cm 3cm, clip=true,width=\columnwidth]{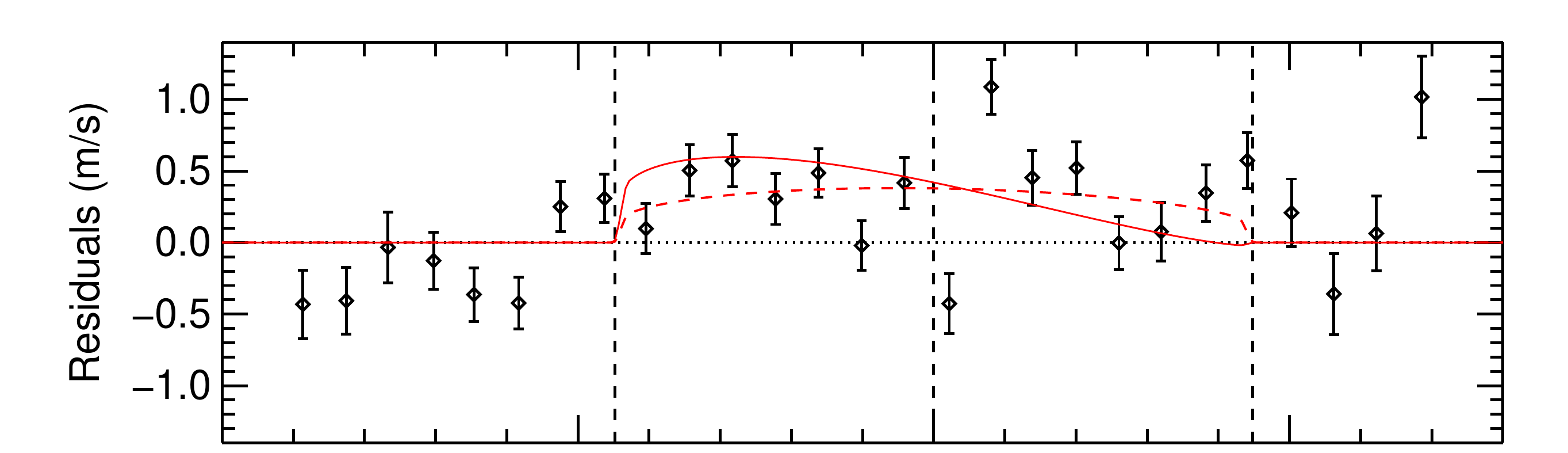}		
\includegraphics[trim=1.5cm 2cm  2.08cm 11.9cm, clip=true,width=\columnwidth]{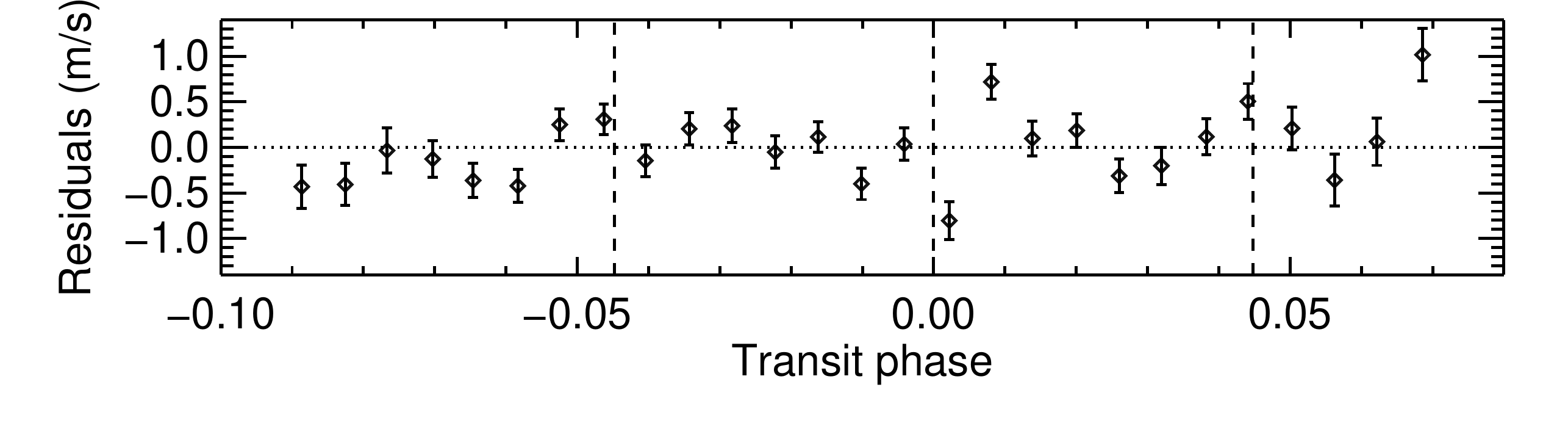}
\caption[]{\textit{Upper panels:} Residuals from the Keplerian fit in dataset E (black diamonds), after the empirical color-effect correction;  (\textit{top}). The dispersion outside the transit is 28\,cm/s. The solid red line shows the best fit of the Rossiter-McLaughlin anomaly with $\lambda$ = 72.4$\stackrel{+9.0}{_{-7.4}}^\circ$. The residuals to the Keplerian + RM fit over the entire sequence yield a dispersion of 28\,cm/s (\textit{bottom}). Vertical dashed lines indicate the times of mid-transit, first, and fourth contacts. \textit{Lower panels:} Residuals from the Keplerian fit after the DRS standard color-effect correction (\textit{top}). The dispersion outside the transit is 43\,cm/s. The dashed red line shows the best fit of the Rossiter-McLaughlin anomaly with $\lambda$ = 88.6$\stackrel{+9.3}{_{-9.9}}$$^\circ$, in agreement with the best-fit obtained after the empirical correction (solid red line as in the upper panels). The DRS-corrected residuals to the Keplerian + RM fit over the entire sequence yield a dispersion of 39\,cm/s (\textit{bottom}). Thus, both the Keplerian fit and the RM fit are better when using the empirical correction rather than the DRS correction.}
\label{fig:residus}
\end{figure}

\begin{figure}[]   % gauche , bas, droite , haut
\includegraphics[trim=4cm 4cm 4cm 4cm, clip=true,width=\columnwidth]{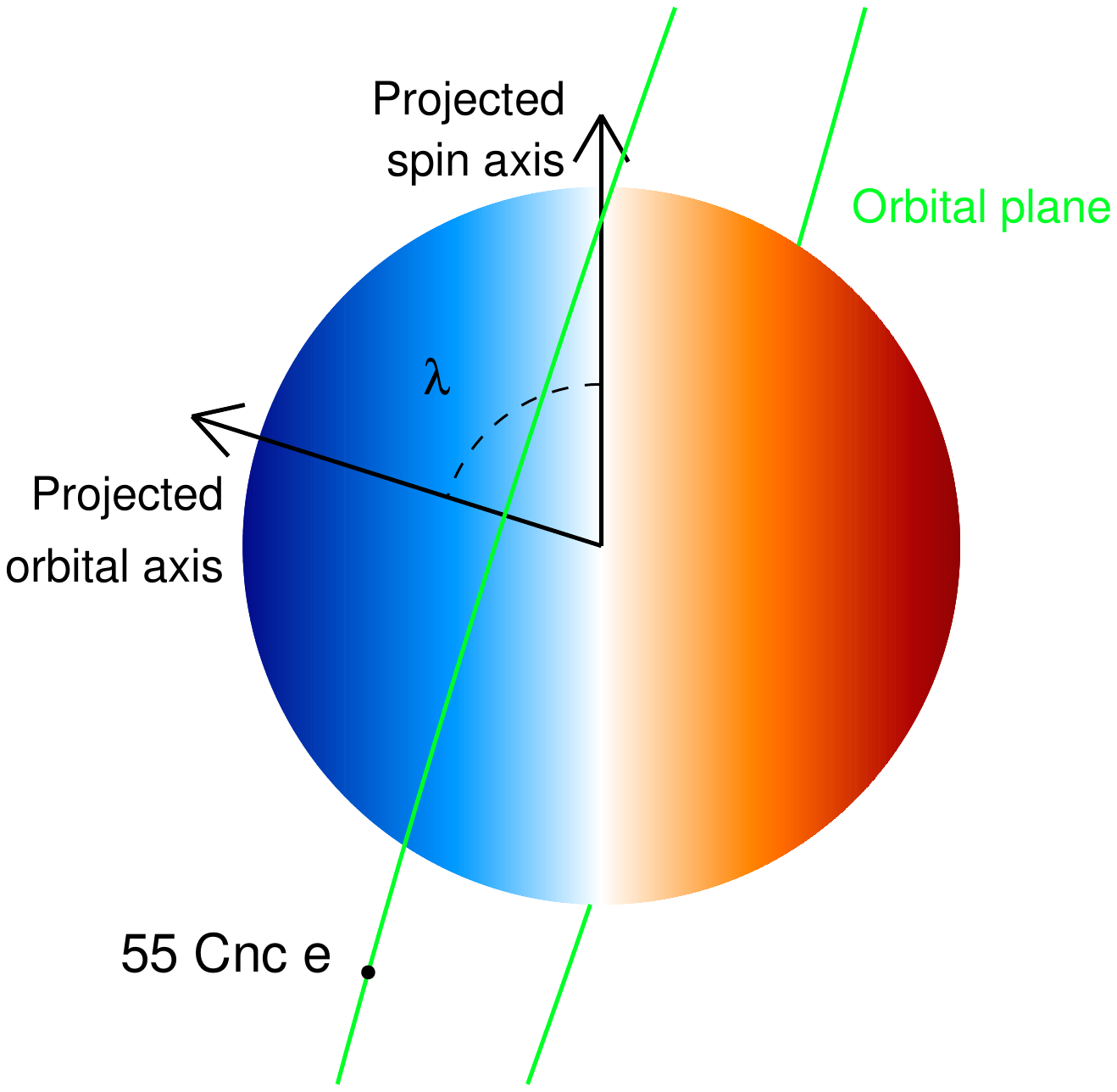}
\caption[]{View of 55 Cnc along the line of sight. With the star rotation, the light emitted by the half of the stellar disk moving toward the observer is blueshifted, while the light from the other half which moves away is redshifted. During the transit, the small Super-Earth 55 Cnc e (shown as a black disk, to scale) transits mainly the blueshifted half of the stellar disk because of its high sky-projected obliquity $\lambda$ = 72.4$^\circ$}
\label{fig:system_view}
\end{figure}

After applying the empirical color-effect correction, radial velocities of dataset E were fitted in order to derive the sky-projected angle $\lambda$ between the planetary orbital axis and the stellar rotation axis. We applied the analytical approach developed by \citet{Ohta2005} to model the form of the Rossiter-McLaughlin anomaly, which is described by ten parameters: the stellar limb-darkening linear coefficient $\epsilon$, the transit parameters $R_\mathrm{p}/R_{*}$, $a_\mathrm{p}/R_{*}$ and $i_\mathrm{p}$, the parameters of the circular orbit ($P$, $T_{0}$, and $K$), the systemic radial velocity $\gamma$, the projected stellar rotation velocity $v$sin$i_{*}$, and $\lambda$. We adopted a linear limb-darkening correction with $\epsilon$=0.648 (\citealt{Dragomir2013}). Parameters for the Keplerian fit are the same as in Sect.~\ref{correc}, and the additional transit parameters for planet 'e' were taken from \citet{Dragomir2013} (see Table~\ref{table:tab_paramsfit}). \\
We computed the $\chi^2$ of the fit on a grid scanning all possible values for $\lambda$, $v$sin$i_{*}$ and $\gamma$. Once the minimum $\chi^2$ and corresponding best values for these parameters were obtained, we calculated their error bars from an analysis of the $\chi^2$ variation. Namely, a given parameter is pegged at various trial values, and for each trial value we run an extra fit, allowing all the other parameters
to vary freely. 1$\sigma$ error bars for the pegged parameter are then obtained when its value yields a $\chi^2$ increase of 1 from the minimum (see, e.g., \citealt{hebrard2002}). We detected the RM anomaly with $\lambda$ = 72.4$\stackrel{+5.0}{_{-6.1}}$$^\circ$ and $v$sin$i_{*}$=3.3$\pm$0.6\,km\,s$^{-1}$ (Fig.~\ref{fig:residus}). The best fit provides a reduced $\chi^2$ of 2.2; to be conservative, we increased the error bars on the radial velocity measurements by a factor $\sqrt{2.2}$ to obtain a reduced $\chi^2$ of 1. As a result, we adopt $\lambda$ = 72.4$\stackrel{+7.4}{_{-9.0}}$$^\circ$ and $v$sin$i_{*}$=3.3$\pm$0.9\,km\,s$^{-1}$. The derived $v$sin$i_{*}$ agrees with the value 2.5$\pm$0.5\,km\,s$^{-1}$ obtained by \citet{vonbraun2011}. The systemic radial velocity $\gamma$ was determined with a particularly high precision of $\pm$8\,cm\,s$^{-1}$. However it depends on the correlation mask, the spectral orders, and the color correction, so the actual barycentric stellar radial velocity is not as accurate. The dispersion of the residuals for the best-fit (28\,cm\,s$^{-1}$) is similar to the estimated out-of transit dispersion (Table~\ref{table:correc}). Results are summarized in Table~\ref{table:tab_paramsfit}.\\
We performed an F-test to evaluate the significance of the RM anomaly detection (see e.g. \citealt{hebrard2011}). We note that in our case the errors may not be normally distributed and independant, and thus the F-test is done as a rough estimate. Including the RM model in the fit improves the $\chi^2$ over the 27 measurements secured during Run E from 95.8 to 51.8, for two additional free parameters ($\lambda$ and $v$sin$i_{*}$). The statistical test indicates there is a probability $>$ 90\% that the $\chi^2$ improvement is due to the RM anomaly detected. We thus conclude that we detected the RM anomaly of 55 Cnc e with $\lambda$ = 72.4$\stackrel{+7.4}{_{-9.0}}$$^\circ$. Its orbit is prograde and highly misaligned, the planet transiting mainly the blueshifted regions of the stellar disk (Fig.~\ref{fig:system_view}).\\

\begin{table*}
%\centering
\caption{Parameters for the Rossiter-McLaughlin analysis of 55 Cnc e.}
\begin{tabular}{lccl}
\hline
\noalign{\smallskip}  
\textbf{Parameter}    & \textbf{Symbol} &  \textbf{Value}  & \textbf{Unit} \\    
\noalign{\smallskip}
\hline
\noalign{\smallskip}
\textbf{Fixed}$^{\dagger}$ & & & \\
Transit epoch 							& $T_{0}$ 							& 2455962.0697$\stackrel{+0.0017}{_{-0.0018}}$ 	& BJD  \\
Orbital period 							& $P$ 									& 0.7365417$\stackrel{+0.0000025}{_{-0.0000028}}$			& day \\
Planet-to-star radii ratio 	& $R_\mathrm{p}/R_{*}$ 	& 0.01936$\stackrel{+0.00079}{_{-0.00075}}$ & \\
Scaled semi-major axis 			& $a_\mathrm{p}/R_{*}$  & 3.523$\stackrel{+0.042}{_{-0.040}}$  & \\
Orbital inclination 				& $i_\mathrm{p}$ 				& 85.4$\stackrel{+2.8}{_{-2.1}}$ & deg\\
Stellar reflex velocity 		& $K$ 									& 6.3$\pm$0.21                   & m\,s$^{-1}$\\
\noalign{\smallskip}
\hline
\noalign{\smallskip}
\textbf{Measured}$^{\ddagger}$ & & & \\
Sky-projected obliquity 							& $\lambda$ 		& 72.4$\stackrel{+12.7}{_{-11.5}}$ & deg\\ 
Projected stellar rotation velocity 	& $v$sin$i_{*}$ &   3.3$\stackrel{+0.9}{_{-0.9}}$ & km\,s$^{-1}$ \\     
Systemic velocity 										&  $\gamma$ 		& 27.40911$\pm$0.00008 & km\,s$^{-1}$ \\
\hline
\multicolumn{4}{l}{$\dagger$: All parameters were taken from \citet{Dragomir2013}, except for $K$ taken from \citet{Endl2012}.} \\
\multicolumn{4}{l}{$\ddagger$: Values are calculated in Section \ref{resultsE}. See also Sect.~\ref{rune_anal} for the uncertainties.}\\
\end{tabular}
\label{table:tab_paramsfit}
\end{table*}

%%%%%%%%%%%%%%%%%%%%%
\section{Validation tests}
\label{verif} 

Because of the small radius of the super-Earth 55 Cnc e and the low rotation velocity of its star, the detection of its RM anomaly is challenging. In addition, we applied a new empirical method for the correction of the color effect. Thus in this section we check the robustness of the detection of 55 Cnc e RM anomaly presented in Sect.~\ref{resultsE}. To be conservative, all error bars on the free model parameters hereafter are scaled and enlarged to maintain a reduced $\chi^2$ of 1.

\subsection{Analysis of Run E}
\label{rune_anal}
In a first time we performed four kinds of tests on dataset E only.

\begin{figure*}
\centering
\begin{minipage}[b]{0.9\textwidth}	
\includegraphics[trim=0cm 1.5cm 0cm 1cm, clip=true,width=\columnwidth]{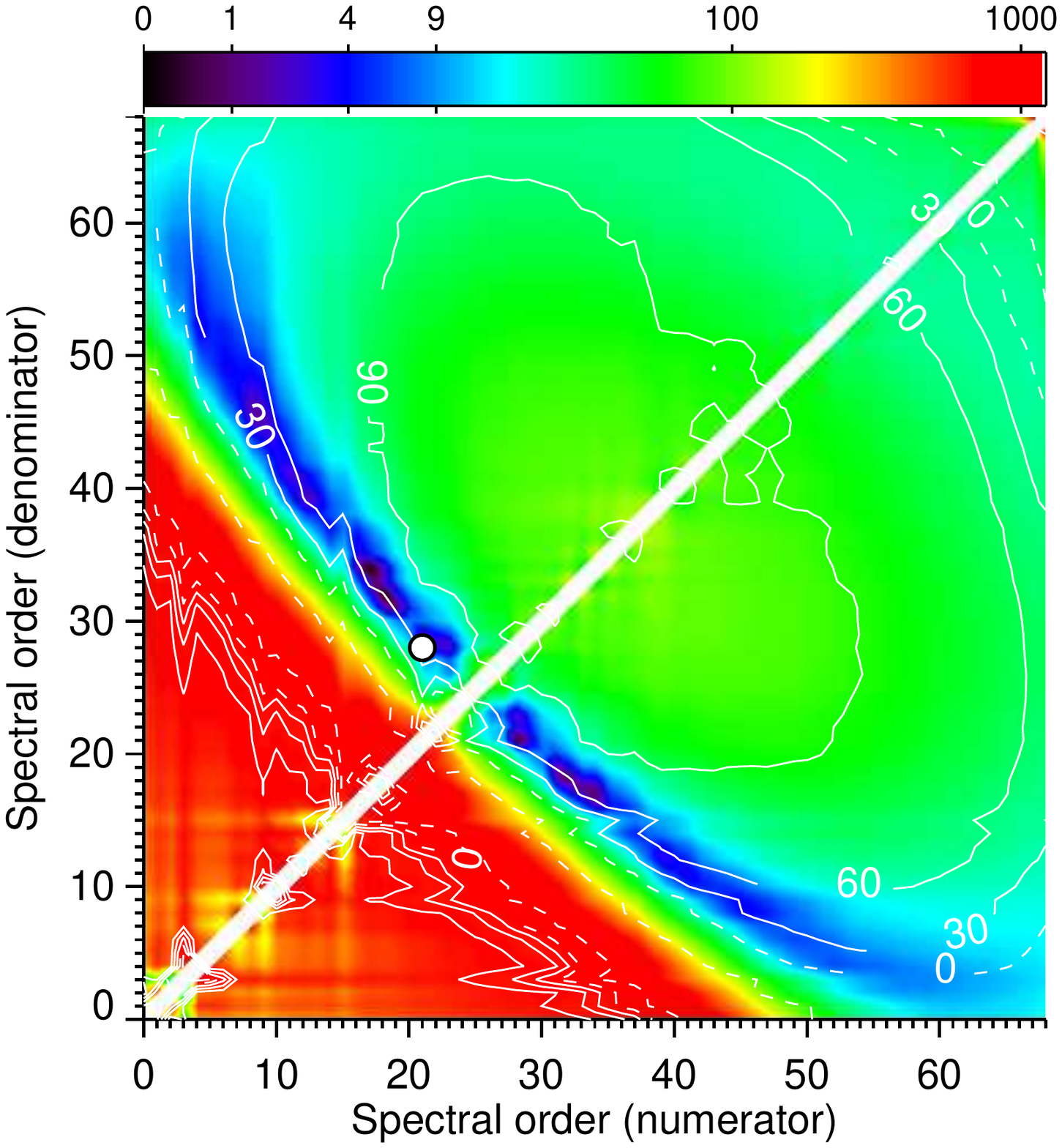}	  % gauche , bas, droite , haut
\end{minipage}	
\caption[]{Dependence of the RM anomaly fit with the spectral orders used to compute the color-effect correction (Run E). Two different spectral orders must be used to quantify the color, which explains the white diagonal line where not fits can be done. White contours show the spin-orbit angles obtained for each color ratio (solid lines for positive values, dashed lines for negative values). The colorscale corresponds to the $\chi^2$ difference with respect to the best fit, obtained with the spectral orders 21 and 28 (white disk) and $\lambda$=72.4$^\circ$. Color ratios in the red part of the diagram show no significant correlation between the residuals of the Keplerian fit and the color ratio. Fits at less than about 3$\sigma$ from the best fit are found in the localized blue area.}
\label{fig:variations_ordres}
\end{figure*}

\begin{enumerate}
	\item \textit{Fitting technique}\\
		%\begin{itemize}
				Minimizing the $\chi^2$ or the out-of-transit dispersion of the residuals, instead of the BIC, gives the same values for the spectral orders used in the empirical color-effect correction provided the polynomial degree is fixed to prevent overfitting.\\
				Calculating the RM model by resampling each 6-minute exposure by 10 (e.g. \citealt{Kipping2010}) has no influence on our results. This was expected because the modelled radial velocity variations during the exposure times are smaller than the error bars (see Fig.~\ref{fig:residus}).\\
		%\end{itemize}
		%\medskip
	\item \textit{Ephemeris}\\
		The empirical color-effect correction is based on measurements outside of 55 Cnc e transit, to prevent removing unintentionnaly its RM anomaly. Thus we investigated how our results depend on 55 Cnc e ephemeris, i.e. the mid-transit $T_\mathrm{0}$, the period $P$, and the transit duration $t_\mathrm{dur}$. We propagated quadratically the errors on the mid-transit time taking into account the number of revolutions accomplished by 55 Cnc e between $T_\mathrm{0}$ and Run E transit epoch. \\
	  %\begin{itemize}
			  Because we do not have enough measurements to constrain the transit mid-time and the period with a high accuracy, we had to choose between the values derived by \citet{Dragomir2013} and \citet{Gillon2012}. We decided to adopt the values obtained by the former as they used long-time baseline MOST photometry of the 55 Cnc sytem, using data obtained in 2012 in addition to the 2011 MOST data used by \citet{Gillon2012}. This reduces the uncertainties on the mid-transit times of our runs, based on their measurement of $T_\mathrm{0}$, and provides a higher-precision value for the orbital period and planet to star radii ratio of 55 Cnc e. Nonetheless we also performed the RM anomaly fit using the ephemeris of \citet{Gillon2012}. In this case $T_\mathrm{0}(E)$, the mid-transit time of run E, is about 11\,mn later, and the transit duration about 8\,mn shorter, than with \citet{Dragomir2013}. Three measurements switch between inside/outside the transit and the color correction is thus different. We obtained $v$sin$i_{*}$ = 3.6$\stackrel{+0.7}{_{-1.0}}$\,km\,s$^{-1}$, and $\lambda$ = 102.9$\stackrel{+11.1}{_{-6.4}}$$^\circ$ at 3$\sigma$ from the previous estimation in Sect.~\ref{resultsE}. It must be noted that the fit is of lower quality, with a reduced $\chi^2$ of 3.6 and a dispersion on the residuals to the RM fit of 36\,cm/s (instead of $\chi^2$=2.2 and a dispersion of 28\,cm/s). Nonetheless we detect again the RM anomaly with a highly misaligned orbit, nearly polar.\\
				We varied $T_\mathrm{0}(E)$ within its 1$\sigma$ error bars using the ephemeris measured by \citet{Dragomir2013}. Because of the large number of revolutions (1065) accomplished by 55 Cnc e since the measure of $T_\mathrm{0}$, the uncertainty on $T_\mathrm{0}(E)$ is about 5.5\,mn, roughly twice that of $T_\mathrm{0}$. We also increased the transit duration by its upper 1$\sigma$ error bar ($\sim$5mn). We found that using lower values for $T_\mathrm{0}(E)$ has no significant impact on our results, whereas with higher values there are not enough measurements after the transit to properly correct for the color effect. This shows that enough measurements must be taken both before and after the transit for the color-effect correction to be efficient.\\
		%\end{itemize}
	  %\medskip
\item \textit{Color-effect correction}\\
		%\begin{itemize}
			  We fitted the RM anomaly to the data extracted with the DRS standard color-effect correction (Fig.~\ref{fig:residus}). The anomaly is detected with $v$sin$i_{*}$ = 2.9$\pm$1.3\,km\,s$^{-1}$ and $\lambda$ = 88.6$\stackrel{+9.3}{_{-9.9}}$$^\circ$. This prograde, highly misaligned orbit is in good agreement with the RM anomaly detected after the empirical color-effect correction, although as expected the quality of the fit is lower with a reduced $\chi^2$ of 3.6 and a dispersion on the residuals to the RM fit of 39\,cm/s.\\
				We investigated how our results depend on the spectral orders used for the empirical color correction. We fixed a linear correction, and calculated the best-fit parameters of the RM anomaly with every possible color ratios. To keep things simple the color ratios are calculated with two spectral orders only (i.e. $i1$=$i2$, $j1$=$j2$). The results are shown in Fig.~\ref{fig:variations_ordres}. As expected the diagram is roughly symmetric, which shows that fits performed with a color ratio or its inverse (e.g. $C_\mathrm{28,28}^\mathrm{21,21}$ and $C_\mathrm{21,21}^\mathrm{28,28}$) have about the same quality and give similar values for $\lambda$. For most ratios $\lambda$ is obtained with a high value between 50 and 110$^\circ$ and stellar velocities between 0.5 and 5\,km\,s$^{-1}$. Only a specific range of spectral orders provides an acceptable adjustment to the data, all with $\lambda$-values around 70$^\circ$. In this range, the best adjustments are obtained with a short separation between the spectral orders at the numerator and the denominator of the color ratio, as expected from our results for all datasets in Table~\ref{table:correc}.\\
		%\end{itemize}
		%\medskip
	\item \textit{Model parameters}\\
		%\begin{itemize}
			The derived values remained within their uncertainties when we varied the limb-darkening coefficient $\epsilon$ between 0.1 and 0.9, as was expected from the precision of our measurements during the ingress and egress.\\
			Increasing the eccentricity of the orbit up of 0.06 (e.g. \citealt{Demory2012}) and using other values for the semi-amplitude of planet 'e' (e.g. \citealt{Nelson2014}) does not change significantly the shape of the Keplerian fit during and around the transit, and hence has little infuence on our results. The same is true for the parameters of the outer planets. \\
		  Varying $R_\mathrm{p}/R_{*}$ and $a_\mathrm{p}/R_{*}$ within their small 1$\sigma$ error bars has no significant influence on our results. Note that we used the values of \citet{Dragomir2013} for these two parameters, as the radius in particular is measured in the optical bandpass of MOST and is more appropriate to our analysis based on HARPS-N data than the radius measured in the infrared with Spitzer (\citealt{Gillon2012}).\\
			We noted that the obliquity is sensitive to the inclination. While the quality of the fit remains unchanged, varying the inclination $i_\mathrm{p}$ between the $1\sigma$ error bars obtained by \citet{Dragomir2013} results in uncertainties of +10.3/-7.2$^\circ$ for $\lambda$. These uncertainties are calculated as the differences between the best-fit values in Sect.~\ref{resultsE} and those obtained while varying $i_\mathrm{p}$. They are similar to the uncertainties derived in Sect.~\ref{resultsE}.\\
			Because of the small amplitude of the measured RM anomaly ($\sim$60\,cm/s), our interpretation of the radial velocity measurements may also be sensitive to the impact of the convective blueshift (CB) effect. We included in our model the calculation of the CB radial velocity blueshift, following the prescription of \citet{shporer2011}. Assuming a solar value for the local convective blueshift (-300\,m\,s$^{-1}$) and a linear limb-darkening law, we found that the CB effect has little influence on our results, with a maximum amplitude of about 4\,cm\,s$^{-1}$ at the center of the transit which is well below the error bars on the radial velocity measurements. Since the distorsion due to the CB effect increases with higher orbital inclinations, we performed again the fit while varying $i_\mathrm{p}$ between its $1\sigma$ error bars. Even then, we obtained similar results as when varying $i_\mathrm{p}$ without CB effect, the derived uncertainties varying by less than 1$^\circ$ for $\lambda$ and 0.1\,km\,s$^{-1}$ for $v$sin$i_{*}$). We also note that 55 Cnc is a G8 star, and thus its local convective blueshift is likely lower than for the Sun.\\
			To conclude, we took a conservative approach to calculate the final error bars on the obliquity and adopted the quadratic sum of the uncertainties obtained in Sect.~\ref{resultsE} and those obtained when taking into account the influence of the inclination to derive $\lambda$ = 72.4$\stackrel{+12.7}{_{-11.5}}^\circ$. We caution that the value of $v$sin$i_{*}$ may not be well constrained by our data, although this has no impact on the obliquity. While the analytic formula derived by \citet{Ohta2005} has been known to underestimate the velocity anomaly (e.g. \citealt{Triaud2009}, \citealt{Hirano2010}), it provides a good approximation when the stellar spin velocity is small enough. Increasing our best-fit value for $v$sin$i_{*}$ by 10\% (i.e. the systematic error in the case of HD\,209458b, which rotates faster than 55 Cnc; \citealt{winn2005}) was found to have no influence on the inferred obliquity. The value of 2.5$\pm$0.5\,km\,s$^{-1}$ obtained by \citet{vonbraun2011} may actually hints that our value for $v$sin$i_{*}$ is overestimated. Assuming $v$sin$i_{*}$ = 2\,km\,s$^{-1}$, we obtained an obliquity of 62$^\circ$ which remains within the derived uncertainties. Smaller $v$sin$i_{*}$ values do not provide a good fit to the data, considering that the RM anomaly is detected with a significant amplitude.\\ 
\end{enumerate}

\subsection{Analysis of all runs}

\begin{figure*}
\centering
\begin{minipage}[b]{0.9\textwidth}	
\includegraphics[trim=0cm 1.88cm 2cm 2cm, clip=true,width=\columnwidth]{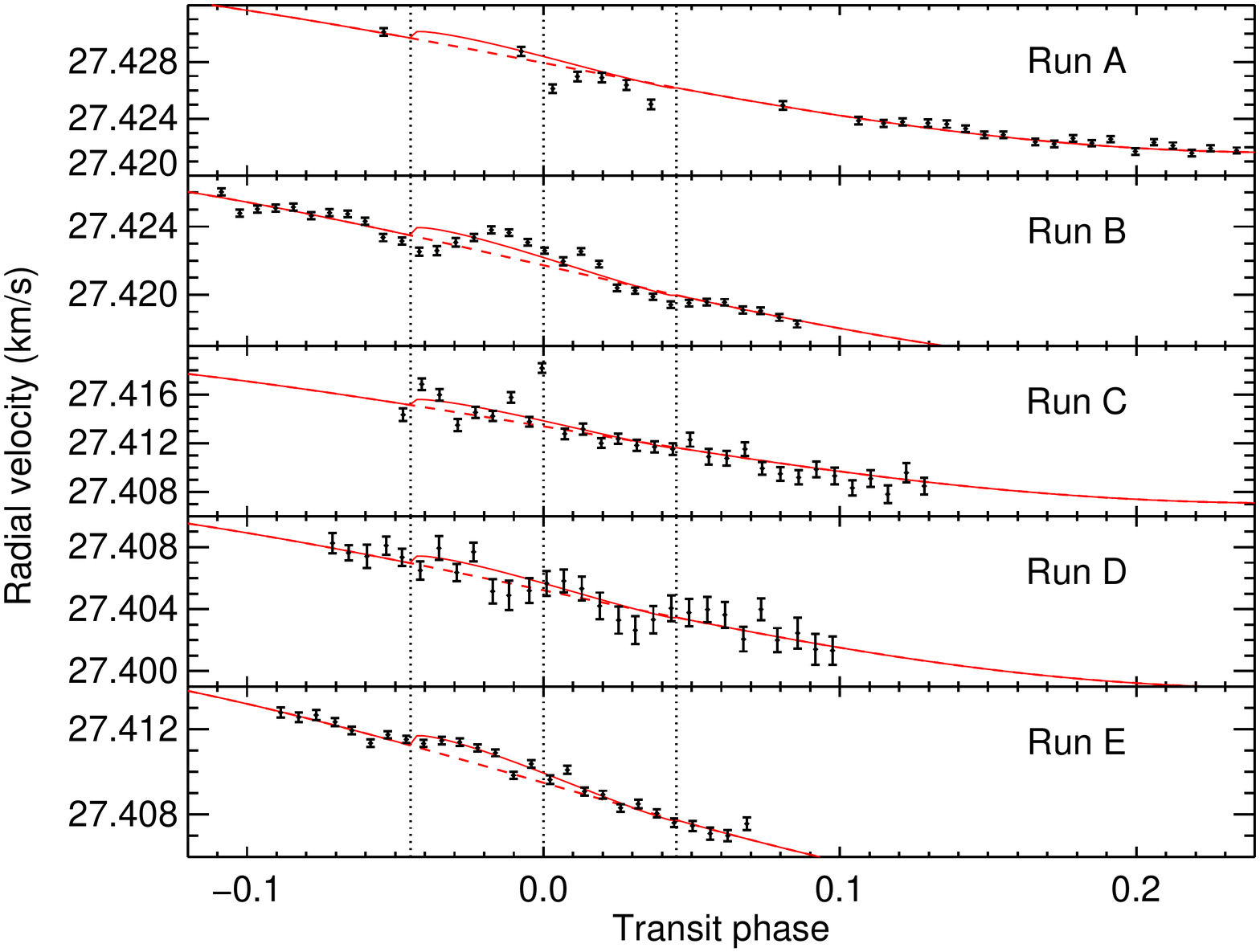}	
\includegraphics[trim=0cm 1cm 2cm 14.25cm, clip=true,width=\columnwidth]{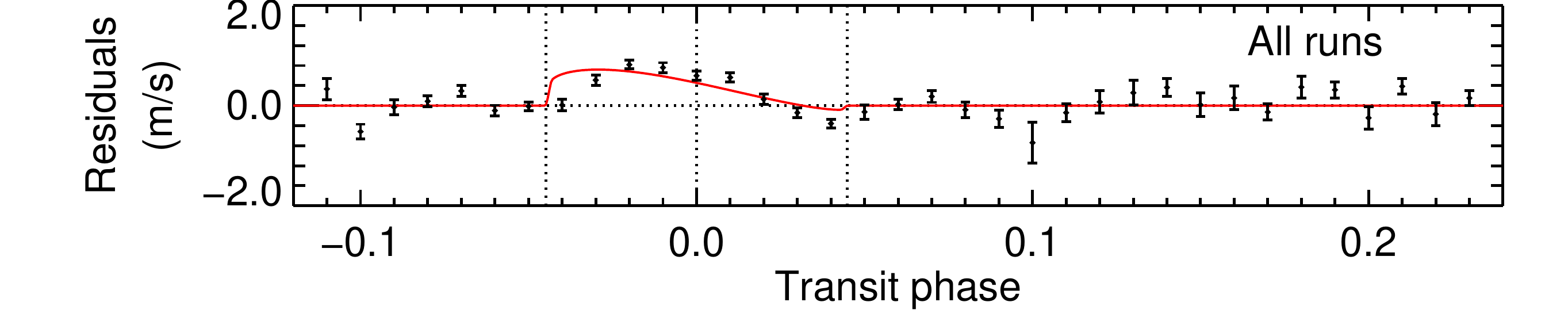}	
\end{minipage}	
\caption[]{Best model of the RM anomaly when fitting datasets A to E simultaneously. Black points show radial velocity measurements as a function of the orbital phase, overlaid with the Keplerian fit ignoring the transit (dashed, red line), and the final fit including the model of the RM anomaly (solid, red line). Vertical dotted lines show the times of mid-transit, first, and fourth contacts. The simultaneous fit to the five runs provides similar results than the fit to Run E only. The bottom panel shows the error-weighted average of the Keplerian residuals over all runs (residuals from the Keplerian fit are first calculated in each run and grouped in common phase bins of width 0.01). Although the combined residuals are dominated by systematic errors, the RM anomaly is clearly visible.} 
\label{fig:kepler_all}
\end{figure*}

Although Runs A to D have lower qualities than Run E (Sect.~\ref{correc}), we checked their consistency with the RM anomaly detected on dataset E. First we fitted simultaneously all datasets (Fig.~\ref{fig:kepler_all}). The error-weighted average of the Keplerian residuals over all runs clearly shows the RM anomaly detection despite the systematic errors (lower panel in Fig.~\ref{fig:kepler_all}). The results are within 1$\sigma$ of those obtained with dataset E only with $\lambda$ = 68.3$\pm$6.6$^\circ$, but the dispersion of the RV residuals to the RM fit is much higher (71\,cm/s). We obtained similar results when fitting all runs simultaneously except run E, with $\lambda$ = 65.2$\pm$8.4$^\circ$ and a dispersion of 77\,cm/s. \\
In a second time we attempted to fit independently each dataset. Run A poorly samples the transit and was observed in two different modes; in addition the data secured during the transit were obtained in a different configuration than the reference data obtained after the transit. This makes Run A suspicious for the RM study, and indeed the fit did not succeed. As mentioned previously (Sect.~\ref{correc}), Run B shows radial velocity oscillations of unclear origin. This may be due to the presence of star spots or granulation on the stellar surface (e.g. \citealt{Boisse2011}; \citealt{dumusque2011}), or an instrumental effect. Despite these perturbations, the empirical color-effect correction allows the detection of the RM anomaly with $\lambda$=77.1$\pm$7.3$^\circ$ but a larger dispersion of the residuals than in run E (54\,cm/s instead of 28\,cm/s). We obtain similar results with dataset C, although the presence of an outlier at orbital phase 0 results in an abnormally large value for $v$sin$i_{*}$. Removing this outlier, we obtain $\lambda$=65.9$\pm$15.2$^\circ$ and a dispersion of 72\,cm/s. Finally we performed an F-test for the RM anomaly in Run D in the same way as in Sect.~\ref{resultsE}, and found a 50\% false positive probability due to the high noise in this dataset, indicating that the anomaly is likely not detected in this run.\\
We conclude that given their lower quality, datasets A to D are in agreement with the RM anomaly detected in Run E.

%%%%%%%%%%%%%%%%%%%%%%%%%%%%%%%%%%%%%%%%%%%%%%%%%%%%%%%%%%%%%%%%%%%%%%%%%%%%%%%%%%%%%%%%%%%%%%%%%%%%%%%%%%%%%%%%%%%%%%%%%%
\section{Discussion}
\label{conclu}

We report the detection of the Rossiter-McLaughlin anomaly of the super-Earth 55 Cnc e, with a sky-projected obliquity $\lambda$ = 72.4$\stackrel{+12.7}{_{-11.5}}^\circ$. The planet is on a prograde and highly misaligned orbit, nearly polar. This detection is mainly based on one high-accuracy transit observed with HARPS-N, and thus more observations of the same quality than Run E are necessary to confirm the detection. 55 Cnc e is the smallest exoplanet for which the projected spin-orbit alignment has been measured \footnote{Stellar obliquities have been measured for the host-stars of three smaller planets (Kepler-50b, 1.71\,$R_{\oplus}$; Kepler-65b, 1.42\,$R_\oplus$; Kepler-65d, 1.52\,$R_\oplus$) using asteroseismology (\citealt{Chaplin2013}), but this technique does not provide a direct measurement of the projected spin-orbit angle}, and is also the planet with the smallest RM anomaly amplitude detected ($\sim$0.6\,m/s) below the Neptune-like exoplanet HAT-P-11 b (1.5\,m/s; \citealt{winn2010b}) and Venus (1\,m/s; \citealt{Molaro2013}). We were able to detect the RM anomaly by devising an empirical color-effect correction for the chromatic variations known to affect radial velocity measurements. This correction is based on the signal-to-noise ratios associated to HARPS-N spectral orders, and it may prove a useful tool to improve the accuracy of RV measurements from other stars and/or instruments. Indeed in the present study our empirical correction was found to improve the dispersion of the RV measurements with respect to the standard DRS correction, and with observing sequences of a few hours we detected the RM anomaly of 55 Cnc e with a high accuracy ($<$30\,cm/s).  \\
The 55 Cnc system is well approximated by a coplanar system (\citealt{kaib2011}; \citealt{Ehrenreich2012}; \citealt{Nelson2014}), and thus all its planets are likely highly misaligned with the stellar spin axis. While most multi-planet systems have been found aligned with the stellar equator, this is the second occurence of a highly misaligned one after Kepler 56 (\citealt{huber2013}). This is a hint that large obliquities are not restricted to isolated hot-Jupiters as a consequence of a dynamical migration scenario. The high obliquity of 55 Cnc e agrees with the fact that lower mass planets are either prograde and aligned, or strongly misaligned (\citealt{hebrard2010,hebrard2011}), although that trend was mainly seen on isolated, Jupiter-mass planets. It is also a new exception to the apparent trend that misaligned planets tend to orbit hot stars (\citet{winn2010a}; the effective temperature of 55 Cnc derived by \citealt{vonbraun2011} is $T_\mathrm{eff}$=5196\,K). The fact that tidal interactions did not align the system (\citealt{barker2009}) during its long lifetime (10.2\,Gy; \citealt{vonbraun2011}) may be due to the low mass of its star, the low mass of its closest companion 55 Cnc e, and the complex dynamical interactions within this compact multiple system (\citealt{Nelson2014}). The particularity of the 55 Cnc and Kepler-56 systems may be the presence of a wide-orbit companion (although such companions may be present in other multiple systems, none have been detected). If the companion is initially inclined with respect to the protoplanetary disk, or with the later inner planets around the primary star, it may misalign their orbital planes while preserving their coplanarity (e.g. \citealt{Batygin2012}; \citealt{kaib2011}). The latter authors investigated this scenario in the case of the 55 Cnc system, whose stellar companion 55 Cnc B was detected at a projected distance of 1065 AU (\citealt{Mugrauer2006}). With a semi-major axis lower than about 4000\,au the gravitationnal influence of 55 Cnc B is strong enough to significantly alter the alignment of the system, providing the star is on a highly eccentric orbit ($e\approxsup$0.95) (\citealt{boue2014b}). \citet{kaib2011} predicted a true obliquity of $\sim$65$^\circ$, which is remarkably consistent with the sky-projected obliquity of 72.4$\stackrel{+12.7}{_{-11.5}}^\circ$ we derived and indicate that the rotation axis of 55 Cnc A is probably not much inclined toward the line of sight. \\

%%%%%%%%%%%%%%%%%%%%%%%%%%%%%%%%%%%%%%%%%%%%%%%%%%%%%%%%%%%%%%%%%%%%%%%%%%%%%%%%%%%%%%%%%%%%%%%%%%%%%%%%%%%%%%%%%%%%%%%%%%

\begin{acknowledgements}
We would like to thank J.-M.~Almenara, F.~Bouchy, M.~Deleuil, R.\,F.~D\'{\i}az, A.~Lecavelier des \'Etangs, G.~Montagnier, C.~Moutou, and A.~Santerne for their help and advice.
This publication is based on observations collected with the 
the HARPS-N spectrograph on the 3.58-m Italian \emph{Telescopio Nazionale Galileo}
(TNG) operated on the island of La Palma by the Fundaci\'on Galileo Galilei of the INAF 
(Instituto Nazionale di Astrofisica) at the Spanish Observatorio del Roque de los Muchachos 
of the Instituto de Astrofisica de Canarias
(programs OPT12B\_13, OPT13B\_30, and OPT14A\_34
from OPTICON common time allocation process for EC 
supported trans-national access to European telescopes).    
We thank the TNG staff for support. The authors acknowledge
the support of the French Agence Nationale de la Recherche
(ANR), under program ANR-12-BS05-0012 ``Exo-Atmos''.
\end{acknowledgements}

\bibliographystyle{aa} % style aa.bst
\bibliography{biblio} % your references Yourfile.bib

\end{document}